\documentclass{osa-article}
\usepackage[english]{babel} 
\usepackage{oe}
\usepackage{epstopdf}
\usepackage{subfigure}  
\usepackage{braket}
\usepackage{cite}
\usepackage{float}
\journal{ome}

\articletype{Research Article}

\begin{document}

\title{Spectral domain inverse design for accelerating nanocomposite metamaterials discovery}

\author{Ashwin K. Boddeti\authormark{1,2}, Abubakr Alabassi\authormark{3}, Vaneet Aggarwal\authormark{1,3}, and Zubin Jacob\authormark{1,2,*}.}

\address{\authormark{1} School of Electrical and Computer Engineering, Purdue University, West Lafayette, IN, 47907, USA\\
\authormark{2}Birck Nanotechnology Center, Purdue University, West Lafayette, IN, 47907, USA\\
\authormark{3}School of Industrial Engineering, Purdue University, West Lafayette, IN, 47906, USA}

\email{\authormark{*}zjacob@purdue.edu\\http://www.electrodynamics.org/} 



\begin{abstract}
Inverse design techniques in the context of nanophotonics have helped in discovery of compact and counter-intuitive structures/shapes. We introduce the concept of spectral domain inverse design to search through the optical trade-space (dispersive permittivity) of nanocomposite  metamaterials. We develop a hybrid optimization technique that combines genetic algorithms and gradient descent methods. We utilize this technique to inverse design an ultra-thin thermophotovoltaic emitter coating material. Our work can lead to an efficient approach to search for new multi-functional optical/thermal metamaterials with desired complex permittivity. 
\end{abstract}

\section{Introduction}

\begin{figure}[!h]
    \centering
    \includegraphics[scale = 0.85]{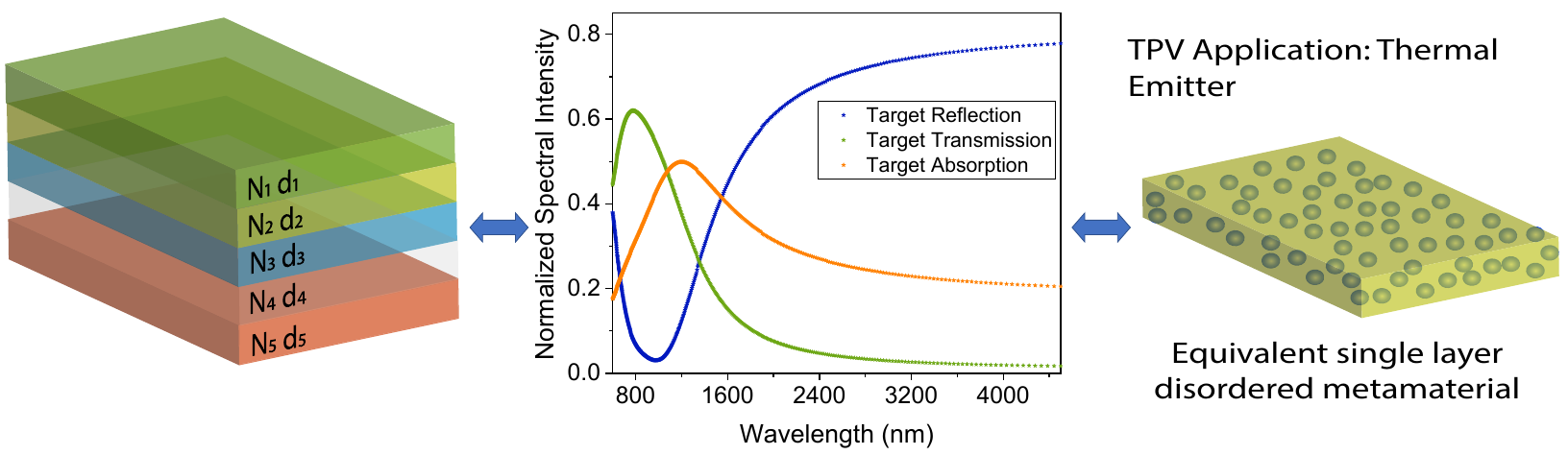}
    \caption{\textbf{Basic Idea:} The spectral features of a N layered multilayer thermal emitter  (TPV emitter) stack are mimicked by a single disordered nanocomposite metamaterial. Using a hybrid optimization technique we find the effective permittivity of this nanocomposite material that as the same reflection and transmission spectral features as the one shown.}
    \label{Fig1}
\end{figure}

Topology optimization, machine learning, gradient descent, needle optimization and genetic algorithms have risen to the forefront of nanophotonics due to their ability to inverse design nanophotonic devices for specific applications\cite{Sean, Chang, Bloom, Sullivan, Tikhonravov, Fan0, Yakovlev, Sigmund1, Sheng, Fan2, JFan, Lu, Dianjing, Malkiel,Lalau , Peurifoy, Liu2, Piggott, Junsuk}. Recent important examples include on-chip wavelength demultiplexer \cite{Piggott}, multilayer thin films\cite{Chang, Bloom, Sullivan, Tikhonravov, Fan0, Dianjing}, chirped-mirrors\cite{Yakovlev}, nanoparticle scattering\cite{Peurifoy}, metasurfaces\cite{JFan, Liu2}, core-shell nanoparticles\cite{Junsuk} to name a few. The common theme in these works is the optimization of structural( topology/shape optimization) \cite{Lalau, Piggott} or configurational (various possible combinations in multilayer thin films)\cite{Fan0, Dianjing} parameters for a given set of available materials to achieve predefined target goals. This motivates the search for analogous techniques that search through the optical dispersion  (dispersive nature of material's permitivitty) trade-space to enable discovery of new nanocomposites and metamaterials.\\\\
Searching through optical material dispersion space is challenging.  Our aim in this work is to put forth spectral domain inverse design  as an approach to achieve desired optical responses through nanocomposite materials and metamaterials. The goal is to search for an optimum material platform with an engineered complex dielectric permittivity that depends on frequency. Thus the target functionality is achieved by changing the effective dielectric permittivity of nano-composites. Coupled with existing structural\cite{Lalau, Piggott} (i.e. shape)  or configurational optimization \cite{Fan0, Dianjing},  our approach can open the route to utilize unique metal-dielectric-semiconductor nanocomposites as well as unconventional emerging 2D materials to realize desired optical responses.
\\\\We perform a proof of principle optimization to show that spectral nanophotonic design can lead to the discovery of new metamaterials in an efficient manner.  Conventional plasmonic thin films of gold, silver, etc. form the starting point for nano-structuring into metamaterials. To discover alternative optimal nanocomposite thin films with desired broadband spectral properties, we utilize effective medium theory in the inverse design algorithm. In our technique, we make use of genetic algorithms and gradient descent techniques in an alternating optimization scheme- thus we call it "Hybrid Optimization". We show that conventional multi-layer thin films can be replaced by an optimized metamaterial nanocomposite Fig.\ref{Fig1}. This hybrid optimization method is used to find an optimal single layer disordered metamaterial for thermalphotovoltaic (TPV) emitter coating applications. 
\\\\

\section{Hybrid Optimization}\label{sec:hybrid}
To start with, here we outline how the hybrid optimization is implemented. First, we specify the desired reflection, transmission and absorption spectra (total of three spectra). In our approach, unlike other implementations \cite{Fan0, Dianjing}, our optimization is highly constrained as we simultaneously optimize on reflection $R^*(\lambda)$, and transmission $T^*(\lambda)$ target spectra. The spectral response of the nanocomposite metamaterial is computed via transfer matrix method. Effective Medium Theory (EMT)\cite{Slovik, Charles-Antoine, Doyle, Malasi} is used to compute the effective permittivity of the nanocomposite metamaterial, as shown in Eq.\ref{EMT}.
\begin{equation}\label{EMT}
    \epsilon_{eff} = \epsilon_{mat} - \frac{\frac{3\rho\epsilon_h}{2x^3}\Big[\sum_{l=1}^{N}i(2l+1)(a_l+b_l)\Big]}{1+\frac{\rho\epsilon_h}{2x^3}\Big[\sum_{l=1}^{N}i(2l+1)(a_l+b_l)\Big] },
\end{equation} where, $x = 2\pi r n_h/\lambda$ is the size parameter ($r$ is the radius, $\epsilon_{h} (n_h) $ is permitivitty (refractive index) of host medium  and $\lambda$ is the wavelength of light), $\epsilon_{mat}$ corresponds to the permittivity of the inclusion material (nanoparticle's material), $l$ is the order of the poles, $a_l$ and $b_l$ are the Mie coefficients that are expressed in terms of Ricatti-Bessel functions. The size parameter sets the cut-off for the number of terms to be considered. It is defined as, $ N = [x+4.05x^\frac{1}{3}+2] $, where the  square brackets mean rounding it to the nearest integer. This enables the  equation  Eq.\ref{EMT} above to handle effects arising due to finite size spherical particle inclusions. This effective medium theory provides the additional degree of freedom in material permitivitty. Here, we use this idea to find an equivalent single layer nanocomposite metamaterial. \\\\
To find the optimum nanocomposite structure, we try to minimize the residuals between the specified target spectra as described by the merit function MF($d,r,\rho,n^1,n^2$). The merit function is defined as Eq.\eqref{MF}, where $d$ is the thickness of the single layer, $r$ is the radius of the nanoparticle inclusions, $\rho$ is the fill fraction, $n^1$ and $n^2$ are the material indices corresponding to a discrete material set. We have used a discrete material set to  account for practical limitations. In this specific problem we have used a total of 22 materials. 
\begin{figure}[ht]
    \centering
    \includegraphics[height = 3.5cm, width = 14cm]{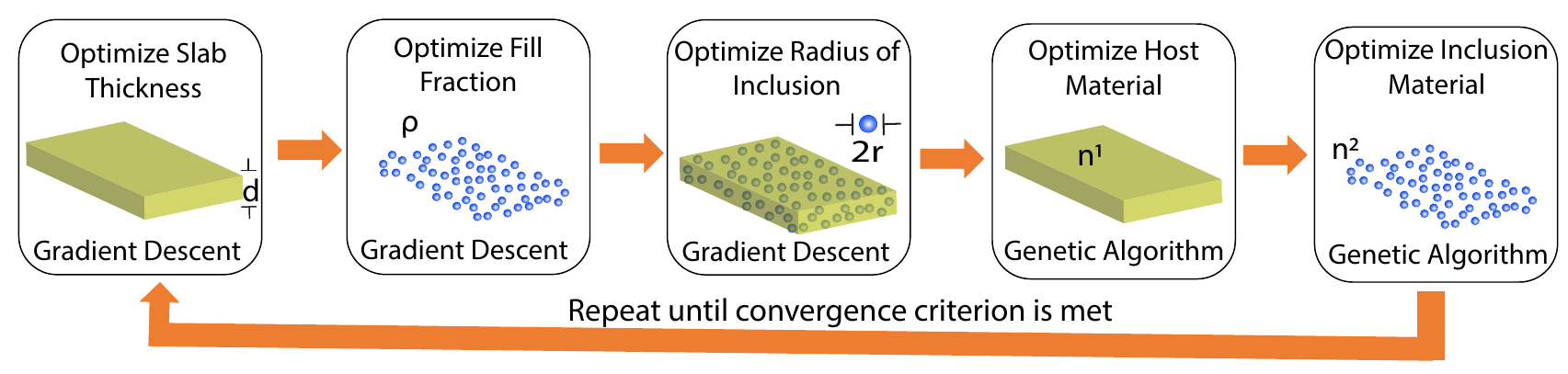}
    \caption{$\textbf{Hybrid Optimization:}$ The flowchart shows the algorithm used to obtain the optimal solution. Gradient descent technique was used for optimizing $d$, $\rho$, $r$, and a discrete search (Genetic algorithm) was used for searching over the material database.}
    \label{Fig2}
\end{figure}
\begin{equation}
\centering 
    MF = \sum_{\lambda} \sqrt{\big[ \big(R(\lambda) - R^*(\lambda)\big)^2 + \big(T(\lambda) - T^*(\lambda)\big)^2 + \big(A(\lambda) - A^*(\lambda)\big)^2  \big]},
    \label{MF}
\end{equation} where $A^*(\lambda)$  is  $1-R^*(\lambda)-T^*(\lambda)$. As we are interested in optimizing over a broad wavelength range, we consider the material dispersion for accurate results. Given the desired target spectra, $R^*(\lambda)$ and $T^*(
 \lambda)$, the optimization problem is formulated as a mixed integer problem,
 \label{Eq.2}
 \begin{equation}
\centering
[d^*,r^*,\rho^*,n^{1*},n^{2*}] = \operatorname*{arg\,min} MF(d,r,\rho,n^1,n^2),
 \end{equation}
where, $d$, $d^*$, $r$, $r^*$, $\rho, \rho^* \in {\rm I\!R^+}$ and $n^1, n^{1*}, n^2, n^{2*} \in {\rm I\!N} $. \\\\Having defined the optimization problem in Eq.\eqref{Eq.2}, we now describe the hybrid optimization approach that we have implemented. One can classify this optimization  as a non-convex optimization problem. The merit function space is very complex and simultaneous optimization of all the parameters did not yield a satisfactory solution (the optimization often converged to a shallow local minimum). As part of the problem considered here has discrete variables, the problem falls in a class of combinatorial optimization. Many problems in this class are NP-hard and efficient optimal algorithms are infeasible unless P=NP \cite{Neumann:2010:BCC:1941919}. NP-hard are class of problems which are at least as hard as the hardest problems in NP where NP is defined as a class of computational problems where given an answer, one can verify the solution in polynomial time by a deterministic Turing machine.  
 \\\\ Considering the complexity of the problem, we have utilized hybrid optimization approach wherein we use a coordinate descent method. 
  A schematic describing the basic algorithmic idea is shown in Fig.\ref{Fig2}. In this method, we optimize one parameter keeping the rest of the parameters fixed. Once the favourable solution for the first parameter is optimized, the next (second) parameter is now optimized using the value of the first optimal parameter as the new initial guess value for the second parameter's optimization run. This process is iteratively repeated across all the parameters until the desired convergence criteria is satisfied. Since each sub-problem of optimizing one parameter at a time converges (decreasing) and the overall problem is bounded from below, the hybrid algorithm converges to a stationary solution\cite{Imre,Bezdek}.
\begin{figure}[ht]
    \centering
     \includegraphics[scale=0.75]{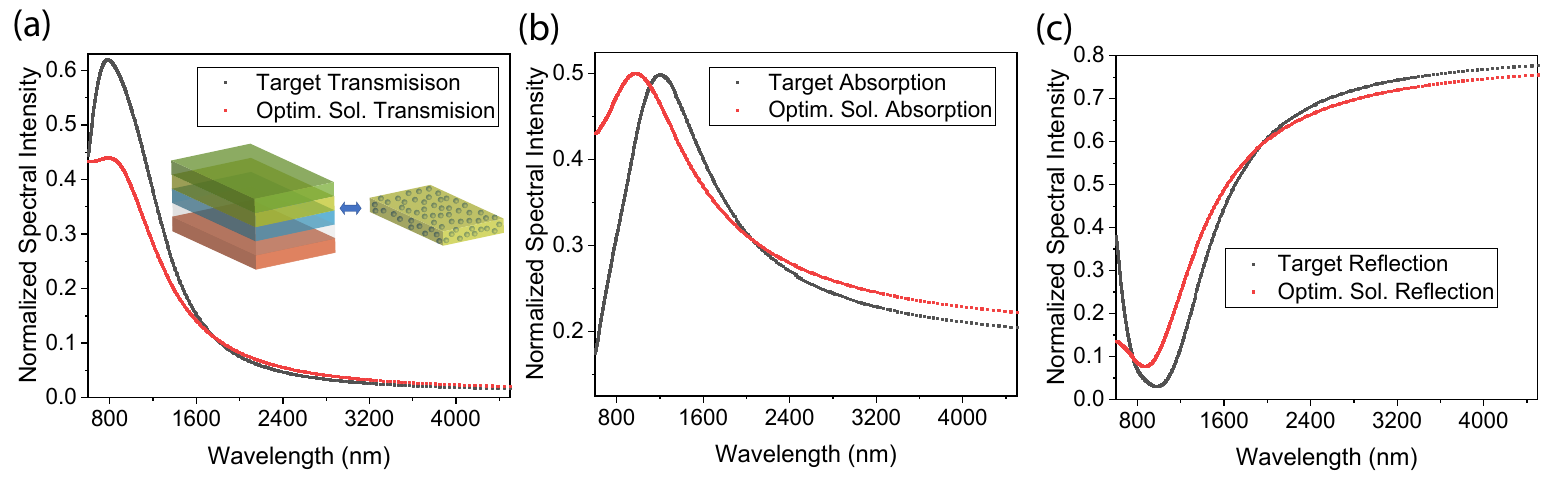}
    \caption{\textbf{Optimal solution:} The figure shows the target spectra and the obtained spectra of single layer disordered metamaterial found using the hybrid optimization technique.The host medium was found to be TiN with HfO2 particle inclusions having a radius of $\sim 34.73nm$ and a fill fraction $\rho \sim 0.84$. Subplots (a), (b) and (c), show the reflection, transmission and  absorption spectra respectively. }
    \label{Fig3}
\end{figure}\\\\
The hybrid approach can primarily be broken down into the following steps: (A) Initialization, (B) Optimal gradient descent, and (C) A discrete optimization. Considering the non-convex nature of the merit function space, it is very important to have an intelligent guess of the initialization parameters. These initialization parameters are chosen by evaluating the merit function value for 10000  sets of five independently generated random numbers. The five random numbers correspond to the thickness (d), fill fraction ($\rho$), radius (r), host material ($n^1$) and inclusion material ($n^2$).  The merit function is evaluated for all these 10000 random sets of values for $(d, r, \rho, n^1, n^2)$. Amongst these, 10 sets having lowest value of the merit function are chosen as the "starting points" (starting guess values) for the hybrid optimization. The hybrid optimization is performed over these 10 sets of initial guess values and the best converged value over these 10 sets is chosen as the output of the algorithm. Using these multiple starting points helps in achieving a better optima. Optimal gradient descent optimization is used for optimizing thickness (d), fill fraction ($\rho$) and radius (r). Active-set constrained optimization algorithm technique \cite{Hager} is used to find the optimal solution. The constraints are based on realistic assumptions set by physical limitations. \\\\Discrete optimization is performed using a Genetic Algorithm (GA) \cite{Whitley, Stefanello}. Genetic Algorithms are a meta-heuristic optimization routine based on the  ideas of natural selection (evolution). Here we utilize this algorithm to search through the material database (22 materials) that make up the nano-composite disordered metamaterial i.e. host material ($n^1$) and inclusion material ($n^2$). The permittivity, $\epsilon(\lambda)$ of each material in the material database is assigned a discrete identification number (unique). The Genetic Algorithm algorithm is utilized to search over these discrete identification numbers to find an optimal material for the host and nanoparticle inclusions. While searching, the corresponding dispersion, $\epsilon(\lambda)$ is used to evaluate Eq. \eqref{Eq.2} and thus the reflection and transmission spectrum using transfer matrix method.
\begin{figure}
    \centering
    \includegraphics[scale=0.23]{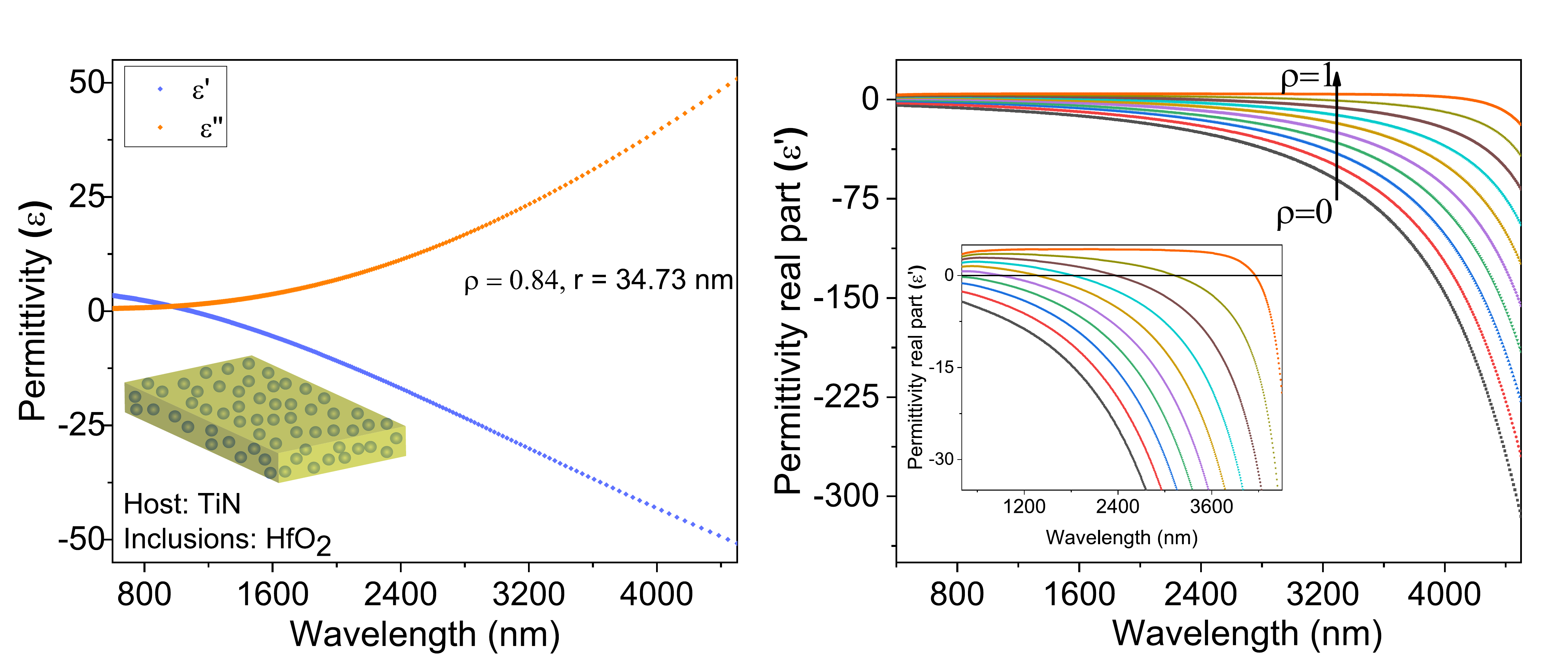}
    \caption{$\textbf{Effective single layer permitivitty and tunning ENZ:}$ The subplot (a) shows the effective permitivitty, $\epsilon = \epsilon' + \iota \epsilon"$ of the single disordered nanocoposite layer. The subplot (b) shows real part of permitivitty $\epsilon'$ calculated for various fill fractions, $\rho$. The inset figure  shows the red shift in the ENZ position with increasing fill fraction, $\rho$ = 0 and $\rho$ = 1 show the bounds within which the ENZ can be varied.}
    \label{Fig4}
\end{figure}
This optimization algorithm is implemented for broad wavelength ranges. The typical run times for the algorithm to reach a reasonable solution is less than $\sim$6 hours. This is very promising considering the highly constrained nature of the optimization problem.
\section{Epsilon-near-zero TPV emitter coating}\label{sec:results}
A thermalphotovoltaic (TPV) system consists of a thermal emitter coating, and a photoconverter (photovolataic (PV) cell ). Selective thermal absorption and emission by the thermal emitter coating can vastly improve the energy conversion efficiency of TPV system. This can be achieved by suppressing the emission of thermal photons below the PV cell bandgap and enhancing it above the cell bandgap \cite{dyachenko2016controlling, Sean2, Sarang}. 

\begin{figure}
    \centering
    \includegraphics[scale =0.7]{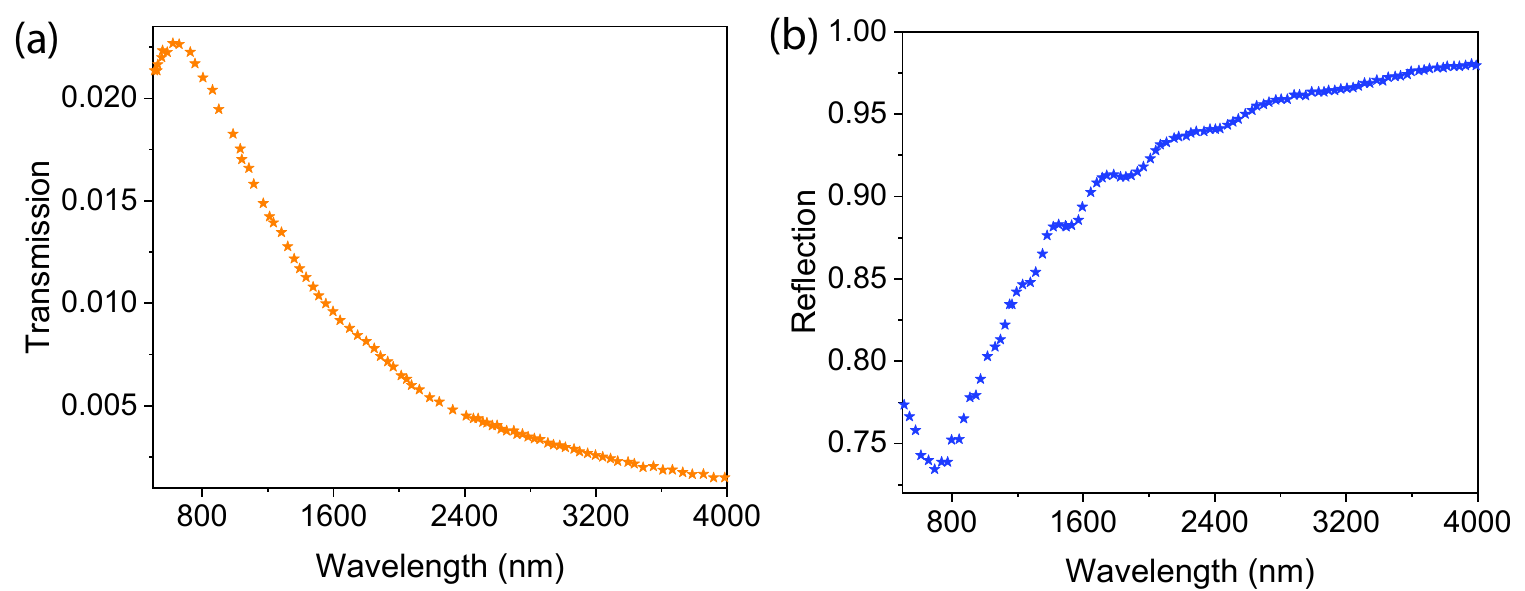}
    \caption{$\textbf{FDTD simulations}$ The subplot (a) shows the transmission spectrum obtained using FDTD simulations. The subplot (b) shows the reflection spectrum obtained using FDTD simulations. This shows qualitative agreement with the EMT. }
    \label{Fig6}
\end{figure}

The hybrid optimization algorithm is used on  realistic target spectra. Reflection and  transmission spectral features of a 5 layered multilayer stack consisting of arbitrary materials and thickness serves as realistic target spectra. The target spectra are shown in Fig.\ref{Fig3} in black. The above introduced hybrid optimization approach is utilized to find the optimal parameters of the single disordered metamaterial layer. Fig.\ref{Fig3} shows the optimal solution obtained utilizing this algorithm. The equivalent single layer solution consists of $TiN$ host and $HfO_2$ particle inclusions.\\\\The effective thickness of this single layer is $\sim 97 nm$ and host particles have a radius of $\sim 34.73 nm$. We note here that this approach is not limited to the infrared spectral region but can also be utilized in the visible wavelegnth range. The effective layer's permitivity is shown in Fig.\ref{Fig4}(a). The versatility of the proposed hybrid optimization is attributed to the ability of tuning the material dispersion properties of the disordered nanocomposite material over a range of wavelengths. 
The spectral position of epsilon near zero (ENZ) of  the nanocomposite layer can be tuned by varying the fill fraction ($\rho$) of the inclusions. This is illustrated in Fig.\ref{Fig4}(b). \\\\We perform $abinitio$ FDTD simulations (Lumerical FDTD solutions) to find the reflection and transmission spectra of the actual disordered metamaterial. A spatial resolution of 1 nm and plane-wave source was used as an excitation source. We found no dependence on the polarization of incident light on the transmission and reflection spectra. Fig.\ref{Fig6} shows the obtained transmission and reflection spectra. There is a qualitative agreement between EMT and FDTD. Low transmission in FDTD results compared to EMT is attributed to the fact that in EMT we assume a uniform, clear medium which is not the case with finite size inclusions. Furthermore, as the higher fill fraction values are not practically  possible to achieve (without the spherical inclusions merging into each other), there is an upper bound on the practically achievable fill fraction. This leads to the spectral mismatch between EMT and FDTD results. However, there is a qualitative agreement between the two. 

\section{Conclusions}\label{sec:concl}
We have introduced the concept of spectral domain inverse design. This is complementary to existing techniques of structural optimization since we search for a complex dielectric permittivity. We have shown an approach to engineer new metamaterials that can be utilized in inverse design problems.  Using this approach we have found a new disordered metamaterial that can be utilized as a TPV thermal emitter.  Our work can lead to a new approach for metamaterials discovery.
\vspace{-.1in}
\section*{Funding}
\vspace{-.1in}
This material is based upon work supported by the U.S. Department of Energy, Office of Basic Energy Sciences under award number DE-SC0017717.
\vspace{-.1in}
\section*{Disclosures}
\vspace{-.1in}
The authors declare no conflicts of interest.

\end{document}